\documentclass[12pt]{iopart}

\usepackage{ae,aecompl}
\usepackage{graphicx}

\usepackage{calc}
\usepackage{upgreek,fixmath}
\usepackage{amssymb,amsbsy,mathrsfs}
\usepackage{color}

\begin{document}

\title{On thermal nucleation of quark matter in compact stars}

\author{B. W. Mintz, E. S. Fraga}
\address{Instituto de F\'\i sica, Universidade Federal do Rio de Janeiro.
Av. Athos da Silveira Ramos, 149.
Centro de Tecnologia - Bloco A.
21941-909. Rio de Janeiro, 
Brazil}

\author{J. Schaffner-Bielich and G. Pagliara}

\address{Institut f\"{u}r Theoretische Physik, Ruprecht-Karls-Universit\"at,
   Philosophenweg 16,  D-69120, Heidelberg, Germany}

\begin{abstract}
     The possibility of a hadron-quark phase transition in extreme astrophysical 
phenomena such as the collapse of a supernova is not discarded by the modern knowledge of
the high-energy nuclear and quark-matter equations of state. Both the density and the
temperature attainable in such extreme processes are possibly high enough to trigger a
chiral phase transition. However, the time scales involved are an important issue.
Even if the physical conditions for the phase transition are favorable (for a system in 
equilibrium), there may not be enough time for the dynamical process of phase conversion 
to be completed. We analyze the relevant time scales for the phase conversion via thermal
nucleation of bubbles of quark matter and compare them to the typical astrophysical time
scale, in order to verify the feasibility of the scenario of hadron-quark phase conversion
during, for example, the core-collapse of a supernova.
\end{abstract}

\maketitle

\section{Introduction}

Extreme densities and temperatures can be achieved during the highly energetic 
core-collapse explosion of a type Ib supernova. In fact, temperatures 
of tens of MeV ($1$MeV$\approx10^{10}$K) and densities as high as $2n_0$ 
($n_0$ is the nuclear saturation density) can be attained right after the bounce, 
when the proto neutron star is still hot and lepton-rich.

In general, SN simulations show that the highest central 
density achieved in the early post-bounce phase is higher than the nuclear 
saturation density $n_0$, but hardly higher than $2n_0$. On 
the other hand, the Equation of State (EoS) for strongly interacting matter 
at low temperature and high baryon density possibly presents a first order 
hadron-quark phase transition at the same density scale. 
Moreover, it has been recently shown \cite{Sagert:2008ka} that if there is 
a hadron-quark phase transition in the early post-bounce phase of a 
core-collapse supernova, it may be decisive for the supernova explosion. 
A natural question that arises is whether the QCD phase transition 
may happen or not during an actual supernova. 
The most important condition is, at least in principle, achievable: 
the critical density for the transition may be lower than the densities 
found during the early post-bounce phase ($\leq2n_0$). However, this condition 
is not enough for the phase transition to happen, once it takes a finite time 
to be initiated and completed, and such time must be shorter than a typical 
time scale of the SN evolution (which we assume to be $\tau_B\lesssim100$ ms) 
\cite{Mintz:2009ay}. This is the central subject of this work. 

By calculating the formation rate of quark matter bubbles at given 
physical conditions, we estimate (or, rather, underestimate) the time scale 
associated with the phase transition and compare it to the duration of the 
pre-deleptonization era in a core-collapse supernova $\tau_B$. This comparison can serve as a 
criterium of feasibility for the formation of quark matter in this scenario, 
because the process of nucleation must occur inside the time window determined 
by $\tau_B$. 


\section{Phenomenological framework}

\subsection{Equations of state}

We discuss the formation of two possible quark phases in the phase transition. In the 
first case, the phase conversion involves two-flavor quark matter 
\footnote{Strange quarks will be produced only later, via weak interaction, as suggested 
in \cite{Bhattacharyya:2006vy}.}. In the second scenario, we consider a fast 
production of strange quarks: since we assume critical densities for the phase 
transition ($n_c$) of the order of two times the saturation density and 
temperatures of a few tens of MeV, it is possible that a small seed of strange matter is 
already present in the system as hyperons and kaons appear 
\cite{Norsen:2002qw,Ishizuka:2008gr}. 
Such particles do not contribute significantly to the pressure 
or to the energy density in the hadronic phase, but their presence 
may trigger a phase transition directly to strange quark matter.

To encompass both possibilities for the quark matter phase, 
we consider two types of high-density EoS, both including the pressure 
from electrons and neutrinos, which are still present at the early post-bounce phase. 
In the first case, we consider only $up$ and $down$ quarks, while in the second we also 
include a massive $strange$ quark. The free parameters of the quark model are the bag constant 
$B$, the mass of the strange quark $m_s$ (when present), and the coefficient $c$ 
\cite{Alford:2004pf}, that effectively accounts for perturbative QCD corrections 
to the free gas pressure \cite{Fraga:2001id}, as follows
(terms ${\cal O}(m_s^4/\mu_s^4)$ were neglected): 
\begin{equation}\label{eq:pressure}
 \hspace{-1.4cm}
p(\{\mu\}) = (1-c)\left[\sum_{i=u,d}\frac{\mu_i^4}{4\pi^2}\right] 
               + (1-c) \frac{\mu_s^4}{4\pi^2} - \frac{3}{4\pi^2}m_s^2\mu_s^2
               + \frac{\mu_e^4}{12\pi^2}+\frac{\mu_\nu^4}{24\pi^2} - B
\end{equation}
For nuclear matter, we adopt the relativistic mean field model 
equation of state with the TM1 parametrization \cite{Shen:1998gq}, 
often used in supernovae simulations.

The equations of state for nuclear matter and quark matter are calculated under conditions 
of local charge neutrality, local lepton fraction conservation (i.e., the two phases have the 
same $Y_L$), and weak equilibrium. Under these assumptions, the conditions of phase 
equilibrium are the equality of the total pressure of 
the two phases $P^H = P^Q$ and condition of chemical equilibrium
$ \mu_n+Y_L \mu_{\nu}^H=\mu_u+2\mu_d+Y_L\mu_{\nu}^Q, $
where $\mu_n$ and $\mu_{\nu}^H$ are the chemical potentials of neutron and neutrinos 
within the nuclear phase, and $\mu_u$, $\mu_d$ and $\mu_{\nu}^Q$ are the chemical 
potentials of up and down quarks and of neutrinos within the quark phase, respectively
\cite{Hempel:2009vp}. 
\footnote{Here we use the zero-temperature equations of state, since a temperature of 
the order of 
a few tens of MeV does not alter considerably the equation of state. For a free 
massless gas, the corrections would be ${\cal O}(T^2/\mu^2)\sim 1\%$.}
Finally, the conditions of local charge neutrality and local conservation of 
the lepton fraction allow us to compute all chemical potentials 
in terms of one independent chemical potential (see, e.g., \cite{Hempel:2009vp}).

\subsection{Thermal homogeneous nucleation}

In first-order phase transitions, if a homogeneous system is brought 
into instability close enough to the coexistence line of the phase 
diagram, its dynamics will be dominated by large-amplitude, 
small-ranged fluctuations, that is, by bubbles of the metastable phase, 
which eventually grow and complete the phase conversion \cite{Gunton_et_al}.

The standard theory of thermal nucleation in one-component metastable systems 
was developed by Langer in the late sixties \cite{Langer:1969bc} (see also 
\cite{Csernai:1992tj}). In this formalism, 
a key quantity for the calculation of the rate of nucleation is the coarse-grained free 
energy functional, that may be cast, in the vicinity of the phase transition, as
\begin{equation}\label{eq:DeltaF}
  \Delta F(R) = 4\pi R^2\sigma - \frac{4\pi}{3}R^3(\Delta p),
\end{equation}
where $\sigma$ is the surface tension of the 
hadron-quark interface. From Eq. (\ref{eq:DeltaF}), we can see that 
$\Delta F(R)$ has a maximum at $R_c \equiv 2\sigma/\Delta p$, 
the critical radius. The bubbles of this size are called {\it critical bubbles} 
and are the smallest bubbles that, once 
formed, can start to drive the phase conversion. 
Therefore, to give a quantitative meaning to the process of nucleation, 
one can calculate the rate $\Gamma$ of critical bubbles created by 
fluctuations per unit volume, per unit time:
\begin{equation}\label{eq:Gamma_def}
 \Gamma = \frac{{\cal P}_0}{2\pi} \,\exp\left[-\frac{\Delta F(R_c)}{T}\right] 
        = T^4 \,\exp\left[-\frac{16\pi}{3}\frac{\sigma^3}{(\Delta p)^2 T}\right],
\end{equation}
where we used (\ref{eq:DeltaF}) and $R_c=2\sigma/\Delta p$.
We make the simple choice 
${\cal P}_0/2\pi=T^4$, which is an overestimate of the actual 
pre-factor \footnote{For a calculation of ${\cal P}_0$ see, e.g., 
\cite{Csernai:1992tj}.}.  
Notice also that the influence of the equation of state is present through 
$\Delta p$. Finally, there is 
a remarkably strong dependence of $\Gamma$ on the surface tension $\sigma$, 
which will be determinant for the nucleation time scale. 

It is convenient to introduce the {\it nucleation time} $\tau$, defined as 
the time it takes for the nucleation of one single critical bubble inside a volume 
of $1km^3$ in the proto-neutron star core: 
$\tau \equiv \left(\frac{1}{1km^3}\right)\frac{1}{\Gamma}$. 
This is the time scale we compare with the duration of the early post-bounce phase
of a supernova event, few hundreds of milliseconds, during which
the formation of quark matter could trigger the supernova explosion. 



\section{Results and discussions}

\subsection{Nucleation times for nonstrange matter}

As our first case, we consider the transition from beta-stable nuclear 
matter to beta-stable quark matter composed of $u$ and $d$ quarks, plus 
electrons and electron neutrinos, with a fixed lepton fraction $Y_L=0.4$ 
and critical baryon density $n_c=1.5n_0$. 

Figure 1 shows the behaviour of the nucleation time of a single 
critical bubble (as defined in the previous section) versus the density, 
in units of $n_0$. 
As expected, the nucleation time $\tau$ has an extremely strong dependence 
on both the density (notice the logarithmic scale for $\tau$) and on the surface 
tension. As discussed in the introduction, we consider that nucleation is 
effective if $\tau <\tau_B\lesssim100$ ms for $n<2n_0$. For low values of 
$\sigma$, nucleation becomes feasible at relatively low densities, although 
the required densities increase steadily as the surface tension rises. 

Next, we compare the previously analyzed $c=0$ case with $c=0.2$, in the case of 
$n_c=1.5n_0$ and $T=20$ MeV, as displayed in Figure 2, where we also show the 
influence of the lepton fraction $Y_L$. 
The contour lines of Figs. 2-4 (on which the nucleation 
time is $\tau=100$ ms) may be considered as ``nucleation thresholds'': given 
a value of surface tension, when the rising density crosses one contour line, 
the nucleation time scale is fast enough for the completion of the phase 
conversion. Notice that the introduction of interactions drastically increases 
the nucleation time, so that only for a low value of the surface tension 
nucleation can be efficient as the density reaches a value close to $2n_0$ 
\footnote{As becomes clear from this analysis, a reliable estimate of $\sigma$ 
for cold dense matter is called for.}. 
Furthermore, notice that given a fixed density, a decrease in $Y_L$ makes 
nucleation more efficient, because deleptonization renders nuclear matter 
less stable.
%
%
\begin{minipage}{210mm}
  \vspace{.2cm}
\begin{minipage}{75mm}
  \hspace{.2cm}
\includegraphics[width=7cm]{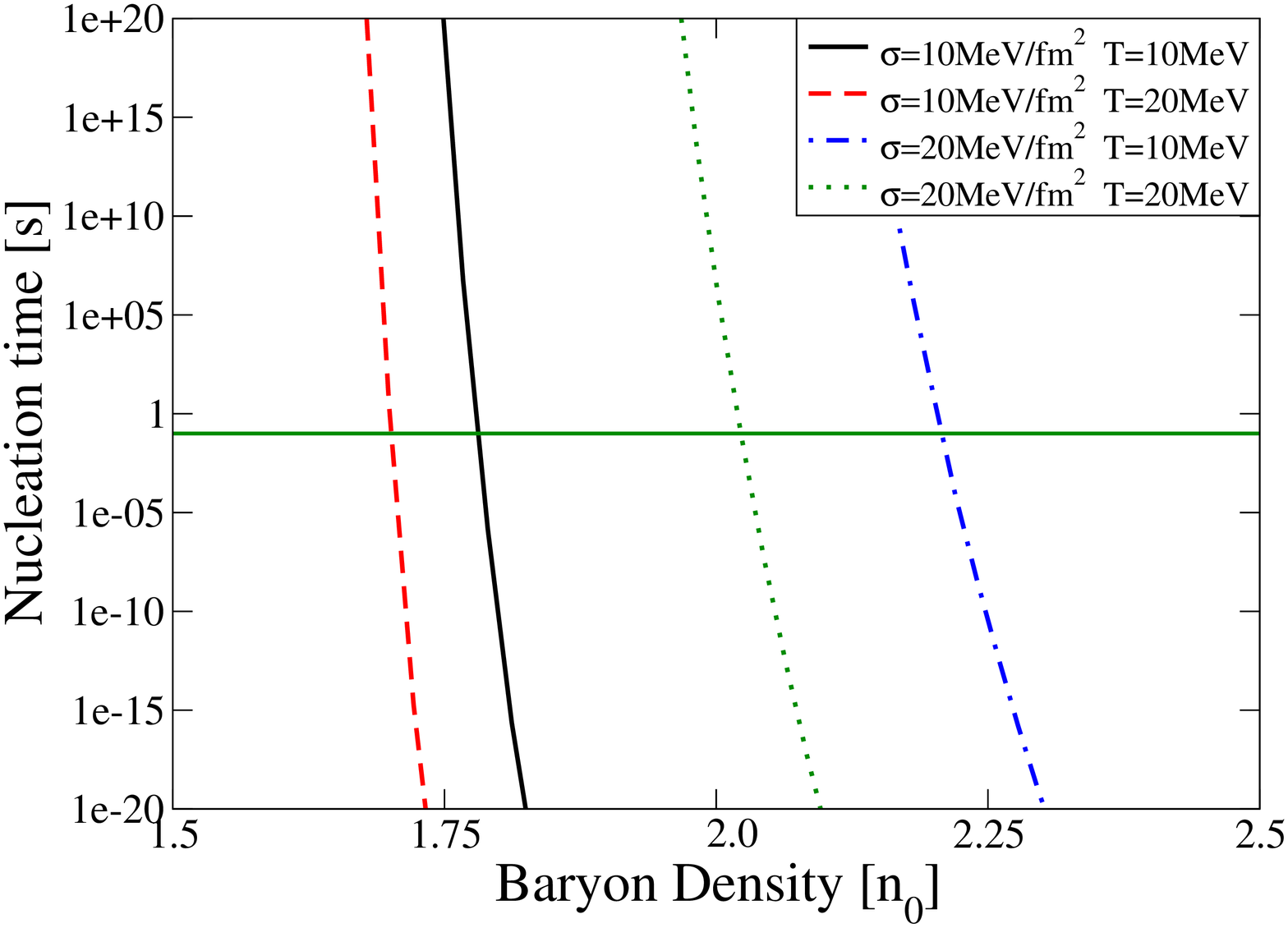}
\vspace{.2cm}
{ \footnotesize{\bf Figure 1.} (From \cite{Mintz:2009ay}.) 
Nucleation time as a function of baryon density for u-d 
quark matter ($n_c=1.5n_0$). The horizontal line corresponds to the nucleation 
threshold $\tau=100$ms.}
\end{minipage}
%
\hspace{.5cm}
\vspace{.25cm}
 \begin{minipage}{75mm}
  \hspace{.2cm}
  \includegraphics[width=7cm,height=5cm]{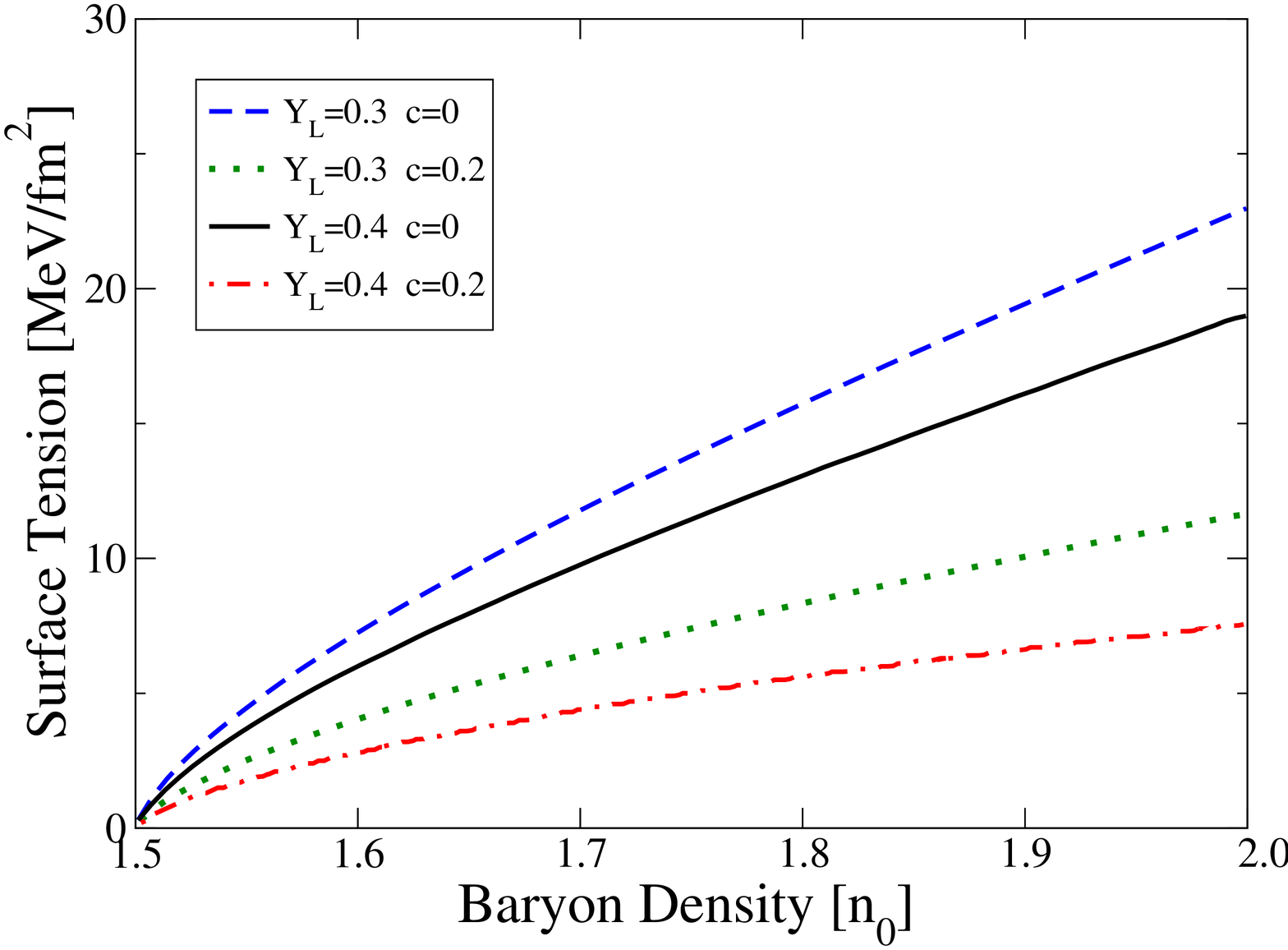}
\vspace{.2cm}

{\footnotesize{\bf Figure 2.} (From \cite{Mintz:2009ay}.)
Lines of constant nucleation time ($\tau=100$ms) 
as a function of $n$ and $\sigma$ with
 noninteracting ($c=0$) and interacting ($c=0.2$) 
u-d quarks, for $Y_L=0.3$ and $Y_L=0.4$.}
\end{minipage}
\end{minipage}


%

\subsection{Nucleation of strange matter}

The introduction of strange quarks makes the EoS stiffer, i.e., for a given baryon 
chemical potential $\mu$ the corresponding pressure becomes higher. Once the 
nuclear EoS is the same, $\Delta p$ will be higher for a given value of $\mu$, 
and the nucleation rate will also be higher. Figure 3 shows 
a comparison between the transition from nuclear matter to either u-d or 
u-d-s quark matter for two values of the lepton fraction $Y_L$ 
($n_c=1.5n_0$, $T=20~$MeV, $c=0$ and $m_s=0$). 
As previously discussed, we can notice that in the 
present case a decrease in $Y_L$ also increases the efficiency of 
thermal nucleation. 



We can also analyze the effect of the strange quark mass on the nucleation time. 
This can be seen in Figure 4, which also shows how the combined effect 
of the strange quark mass and of (perturbative) strong interactions can strongly 
increase the nucleation time ($n_c=1.5n_0$, $T=20~$MeV and $Y_L=0.4$). 
Notice that the presence of interactions among quarks suggest that 
the value of $\sigma$ should not exceed $\sim 20$ MeV/fm$^2$ (with the 
realistic value $m_s=100$ MeV), if we require nucleation to be efficient.
\begin{minipage}{210mm}
%
\begin{minipage}{75mm}
  \hspace{.2cm}
\includegraphics[width=7cm]{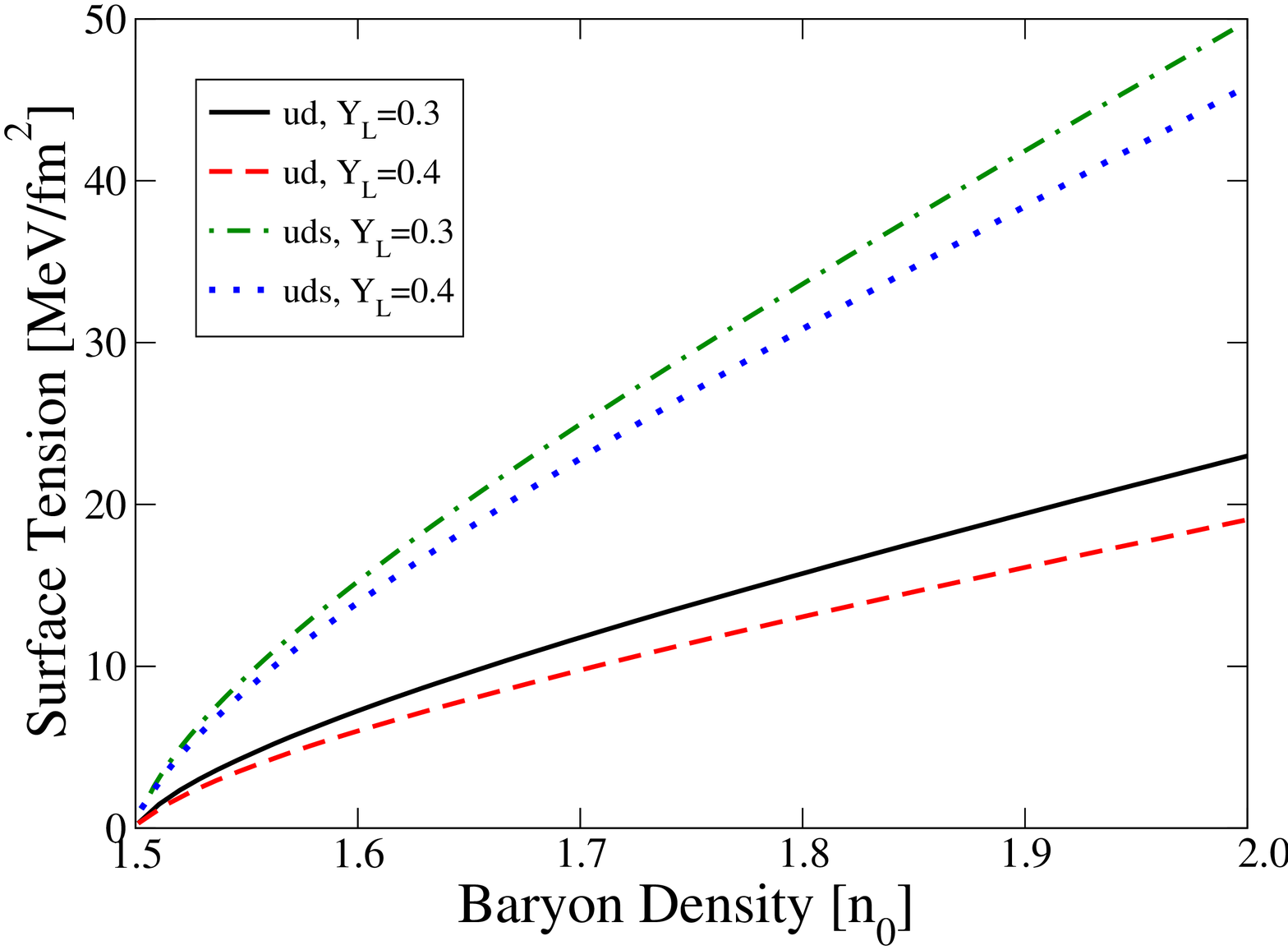}
\vspace{.2cm}
{ \footnotesize{\bf Figure 3.} (From \cite{Mintz:2009ay}.)
Lines of $\tau=100~$ms for 
the transition to u-d or u-d-s quark matter, 
with lepton fraction $Y_L=0.3,\,0.4$.}
\end{minipage}
%
\hspace{.5cm}
%
 \begin{minipage}{75mm}
  \hspace{.2cm}
  \includegraphics[width=7cm]{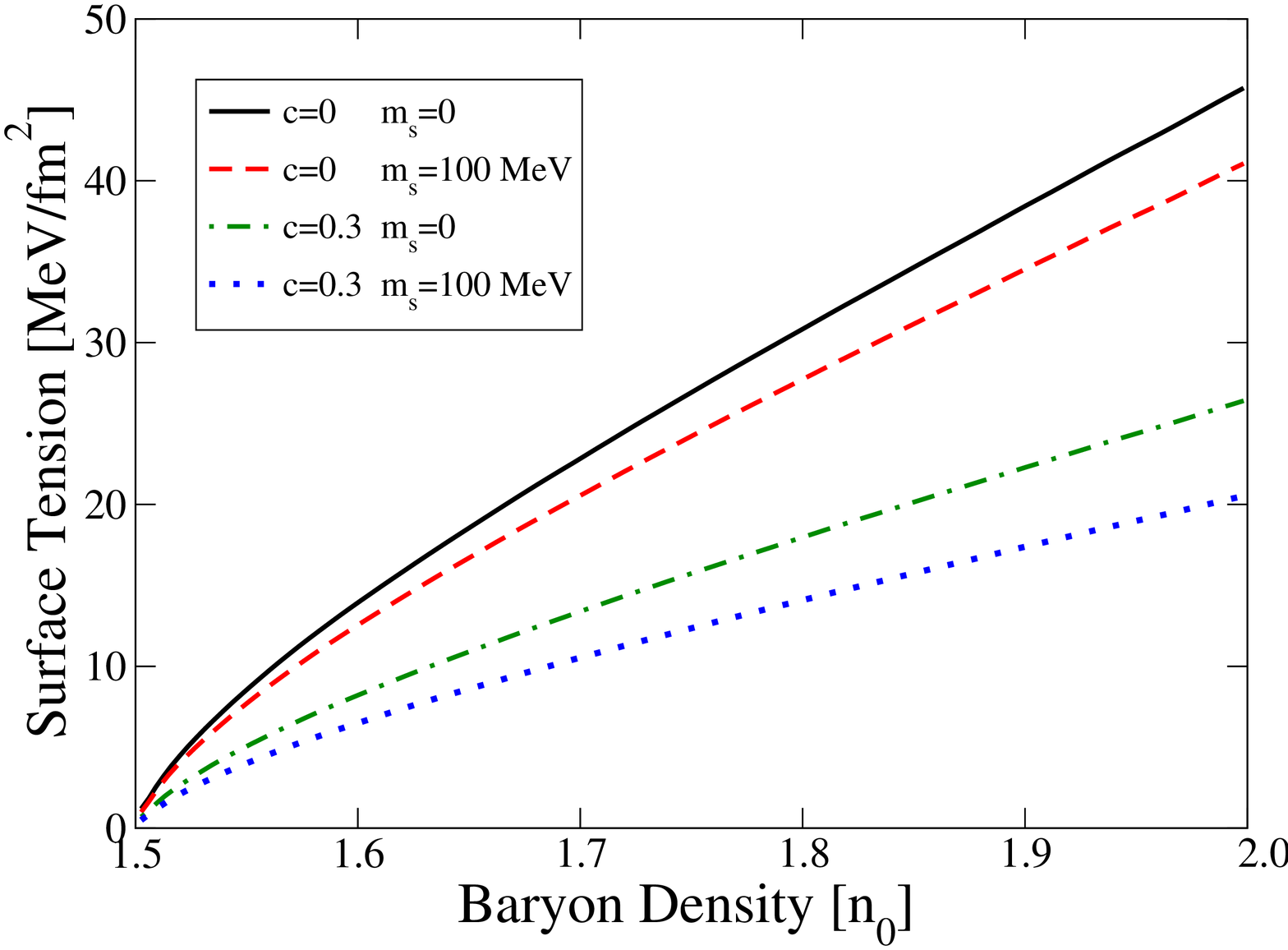}
\vspace{.2cm}
{ \footnotesize{\bf Figure 4.} (From \cite{Mintz:2009ay}.)
Lines of $\tau=100~$ms for 
$c=0$ and $c=0.3$, and for quark mass $m_s=0$ and $m_s=100$ MeV.}
\end{minipage}
\end{minipage}

\section{Conclusions}

We investigated the possibility of formation of quark matter in
supernova matter, i.e., for temperatures of the order of a few tens of
MeV and in the presence of trapped neutrinos, assuming that the
corresponding critical density does not exceed $2 n_0$. By calculating 
the nucleation rate for different values of the free
parameters, we argued
that thermal nucleation of droplets of the quark phase is possibly the
dominant mechanism for the formation of the new phase. 

As expected from Eq. (\ref{eq:DeltaF}), the surface tension is 
the physical quantity which mainly controls the nucleation process and,
within our choices of physical conditions (for the EoS we tested), the 
value of $\sigma$ should not exceed $\sim 20~$MeV/fm$^2$
if the hadron-quark phase transition in core-collapse supernovae 
occurs via thermal nucleation. 


\section*{Acknowledgments}

B.~W.~M. and E.~S.~F. thank CAPES, CNPq, FAPERJ and FUJB/UFRJ for financial support.  
The work of G.~P. is supported by the
Alliance Program of the Helmholtz Association (HA216/EMMI) and by the Deutsche
Forschungsgemeinschaft (DFG) under Grant No. PA1780/2-1.
J.~S.~B. is supported by the DFG through
the Heidelberg Graduate School of Fundamental Physics. The authors also thank
the CompStar program of the European Science Foundation.
      

\section*{References}


\end{document}